\documentclass[12pt,epsfig,a4]{article} 
\newcommand{\vs}{\vspace{-0.25cm}}
\usepackage{amsmath,graphicx}
\begin{document} 
\begin{center}
{\Large{\bf Reducible chiral four-body interactions in nuclear 
matter}\footnote{This work 
has been supported in part by DFG and NSFC (CRC110).}  }  

%\medskip

 N. Kaiser and R. Milkus \\
\medskip
{\small Physik-Department T39, Technische Universit\"{a}t M\"{u}nchen,
   D-85747 Garching, Germany\\

\smallskip

{\it email: nkaiser@ph.tum.de}}
\end{center}
\medskip
\begin{abstract}
The method of unitary transformations generates five classes of leading-order 
reducible chiral four-nucleon interactions which involve pion-exchanges and a 
spin-spin contact-term. Their first-order contributions to the energy per particle of 
isospin-symmetric nuclear matter and pure neutron matter are evaluated in detail. 
For most of the closed four-loop diagrams the occurring integrals over four 
Fermi-spheres can be reduced to easily manageable one- or two-parameter integrals. 
One observes substantial cancelations among the different contributions arising from 
2-ring and 1-ring diagrams. Altogether, the net attraction generated by the chiral 
four-nucleon interaction does not exceed values of $-1.3$\,MeV for densities 
$\rho<2\rho_0$. 
\end{abstract}

\section{Introduction and summary}
According to their modern description and construction in chiral effective field
theory, nuclear forces are organized in a hierarchical way \cite{evgeni,hammer}. For 
generic few- and many-body observables the contributions arising from two-nucleon 
interactions are larger than those from three-nucleon forces, and the latter are 
again more important than possible  corrections due to four-body forces. After a
renormalization-group evolution to lower resolution scale such chiral low-momentum 
interactions exhibit desirable convergence properties in 
perturbative calculations of many-nucleon systems and infinite nuclear matter 
\cite{vlowkreview,achimnuc,3bodycalc,hebeler}. In particular, the inclusion of the 
leading-order chiral three-nucleon interaction (consisting of contact, 
$1\pi$-exchange, and $2\pi$-exchange components) is essential in order to achieve 
saturation of isospin-symmetric nuclear matter in this framework. The calculation of 
the sub-leading chiral three-nucleon interaction, built up by many pion-loop diagrams 
etc., has been completed in ref.\,\cite{3bodyn3lo}. By employing chiral two-, three-, 
and four-nucleon interactions up to order N$^3$LO the equation 
of state of neutron matter at zero temperature has been studied in detail by the 
Darmstadt group \cite{achim,darmstadt}. A good convergence pattern has been observed 
for the second and third order particle-particle diagrams with two-nucleon 
potentials only and when including additionally three-nucleon forces in the form of 
density-dependent two-body interactions. Moreover, in ref.\,\cite{darmstadt} the 
N$^3$LO three-body and four-body contributions to the energy per particle of 
isospin-symmetric nuclear matter have been estimated and large effects from the 
sub-leading chiral three-nucleon force have been found together with small 
corrections from the leading-order chiral four-nucleon interactions.

An analytical calculation of chiral four-body interactions in nuclear matter and pure
neutron matter has been performed recently in ref.\,\cite{4body}. The long-range 
four-nucleon interaction is generated by pion-exchanges that are combined by the 
chiral $4\pi$-vertex (off-shell $\pi\pi$-interaction) or the chiral 
$N\!N3\pi$-vertex. The contributions arising from the 2-ring and 1-ring diagrams 
related to this ''irreducible'' four-nucleon interaction turned out to be very small, 
with values below $0.6\,$MeV for densities $\rho<0.4\,$fm$^{-3}$. However, when 
including the $\Delta(1232)$-isobar as an explicit degree of freedom and counting 
the delta-nucleon mass splitting $\Delta = 293$\,MeV as a small scale (comparable to 
the pion mass $m_\pi$ or the Fermi momentum $k_f$) a new class of leading-order 
long-range four-nucleon interactions arises. These are mediated by two-fold  
$\Delta(1232)$-isobar excitation of nucleons and the exchange of three pions. The 
analytical calculation of the pertinent 3-ring, 2-ring, and 1-ring diagrams at 
four-loop order has lead to a moderately repulsive contribution of $\bar E(\rho_0)
=2.35\,$MeV at nuclear matter saturation density $\rho_0 = 0.16\,$fm$^{-3}$. However, 
the curve for $\bar E(\rho)$ rises strongly with the density, reaching about 10 
times that value at $\rho= 2\rho_0$. In the case of pure neutron matter the same 
$\Delta(1232)$-induced four-neutron interactions lead to a repulsive contribution 
that is about half as strong (see Fig.\,13 in ref.\,\cite{4body}). 

The complete set of leading-order four-nucleon interactions in chiral effective 
field theory has been constructed by Epelbaum in refs.\,\cite{evgeni4na,evgeni4nb}
using the method of unitary transformations. This method allows to project the 
dynamics of the interacting pion-nucleon system into the purely nucleonic subspace 
relevant for few-nucleon systems at energies below the pion-production threshold. 
In addition to the ''irreducible'' chiral four-nucleon interaction mentioned above, 
which follows (via the Feynman diagram technique) directly from the  vertices of the 
chiral Lagrangian ${\cal L}_{\pi N}^{(1)}+ {\cal L}_{\pi\pi}^{(2)}$, the method of unitary 
transformations gives rise to eight classes of ''reducible'' four-nucleon diagrams. 
In ref.\,\cite{evgeni4nb} it is shown that disconnected diagrams always lead to 
vanishing $4N$-forces and with the kinematically suppressed pion-energies at a 
Weinberg-Tomozawa  $\pi\pi N\!N$-vertex one obtains that actually three classes (III, 
VI, and VIII) vanish at leading order. The nonvanishing $4N$-interactions, grouped 
into the classes I, II, IV, V, and VII, involve multiple pion-exchanges and a 
spin-spin contact-interaction with coupling constant $C_T$. The associated isoscalar 
central $N\!N$-contact term proportional $C_S$ is absent because it drops out of the
commutators arising in the construction of the reducible $4N$-interaction. Note 
that the ``irreducible'' four-nucleon interaction, which constitutes a part of class 
II, is called $V^e+V^f$ in ref.\,\cite{evgeni4na}. Moreover, one finds that in an 
anti-symmetrized four-neutron state only the (diagonal) matrix-element of $V^a$ 
(class I) is nonvanishing and an explicit expression has been given for it in 
eq.(A14) of ref.\cite{darmstadt}. Analogous expressions for the four-nucleon 
matrix-elements of the other classes have so far not been published.

The purpose present paper is to evaluate analytically the five classes (VII, V, IV, 
II, I) of reducible chiral four-nucleon interactions ($V^n, V^l, V^k, V^c, V^a$ in 
the notation of ref.\,\cite{evgeni4na}) in isospin-symmetric nuclear matter and 
pure neutron matter. In order to facilitate a uniform treatment we will first
introduce in section\,2 a general representation of $4N$-interactions in terms of a 
product of four (effective) single-nucleon vertices and three ''propagator''-factors.
The latter will be ordinary pion-propagators, their squares, or just constants. The 
first-order ''Hartree-Fock'' contribution to the energy density of nuclear matter is 
obtained from the connected four-body diagram by closing its four nucleon-lines in 
all possible ways. The  actually occurring chiral vertices have the convenient 
property that closing any nucleon-line to itself leads to a vanishing spin-trace or 
isospin-trace. Therefore, one needs to consider here only the closed 2-ring and 
1-ring diagrams generated by the reducible chiral $4N$-interactions. These four-loop 
diagrams are then evaluated individually for the five different classes (in the order 
VII, V, IV, II, I) in the subsections 2.1 up to 2.5. In most cases the occurring 
integral over the product of four Fermi spheres of radius $k_f$ can be reduced to an 
easily manageable one- or two-parameter integral, that depends on the dimensionless 
ratio $k_f/m_\pi$. The 1-ring diagrams with crossed pion-exchanges related to class 
II and I allow only for a partial analytical reduction, such that six-parameter 
integrals remain. The numerical results for the energy per particle $\bar E(\rho)$ 
as a function of the nucleon density $\rho=2k_f^3/3\pi^2$ are presented and discussed 
in section 3. We consider two opposite values of the spin-spin contact-coupling, 
$C_T=0.22\,$fm$^2$ and  $C_T=-0.45\,$fm$^2$, taken from table I in ref.\,\cite{
darmstadt}. One observes substantial cancelations between the contributions from
 different classes. In particular, the $C_T$-independent classes II and I provide 
the largest repulsive and attractive contribution, respectively. Altogether, the net 
attraction generated by the reducible chiral four-nucleon interaction stays above 
values of $-1.3\,$MeV for densities $\rho<2\rho_0=0.32$\,fm$^{-3}$. This feature 
holds for the whole range $|C_T|<0.5\,$fm$^2$ of the spin-spin contact-coupling 
$C_T$. In the case of pure neutron matter, where only class I contributes, the 
result for the energy per particle $\bar E_n(\rho_n)$ is determined entirely by a 
2-ring diagram, since the three types of 1-ring diagrams compensate each other almost 
completely. The corresponding attractive energy per particle goes approximately as 
$\rho_n^{7/3}$ and it reaches a value of about $-2.1$\,MeV at $\rho_n=0.4$\,fm$^{-3}$.  

In summary one finds that the chiral four-nucleon correlations studied in this work 
are at least one order of magnitude smaller than those provided by the strongly 
coupled $\pi N\Delta$-system \cite{4body} with its small mass-gap of $293$\,MeV.

\section{Diagrammatic calculation of four-body interactions in nuclear 
matter}
In this section we derive analytical expressions for the first-order contributions 
to the energy per particle of nuclear matter as they arise from the reducible chiral 
four-nucleon interactions of ref.\,\cite{evgeni4nb}. In order to facilitate a 
uniform treatment of the five different classes (VII, V, IV, II, I) we introduce 
first a generic representation of $4N$-interactions in terms of a product of four 
(effective) single-nucleon vertices and three ''propagator''-factors. The latter 
will be ordinary pion-propagators, their squares, or just constants. Fig.\,1 shows 
the intended generic form of the reducible chiral four-nucleon interaction, 
represented as an on-shell scattering process $N_1N_2N_3N_4\to N'_1N'_2N'_3N'_4$. The 
labels $a,b,c$  denote isospin-indices of exchanged pions and $i,j$ denote 
components of (spin)-vectors. These vector-indices are introduced by the spin-spin 
contact-term proportional to $C_T$, and if present, the left and/or right 
''propagator''-factor is a constant. The momentum transfers $\vec q_a=\vec p_1\,\!\!'-
\vec p_1$, $\vec q_c=\vec p_4\,\!\!'-\vec p_4$ and $\vec q_b = \vec p_2\,\!\!'-
\vec p_2+\vec q_a= \vec p_3 -\vec p_3\,\!\!'-\vec q_c$ are given by differences of 
out- and in-going nucleon momenta.  
\begin{figure}[t!]
\begin{center}
\includegraphics[scale=0.6,clip]{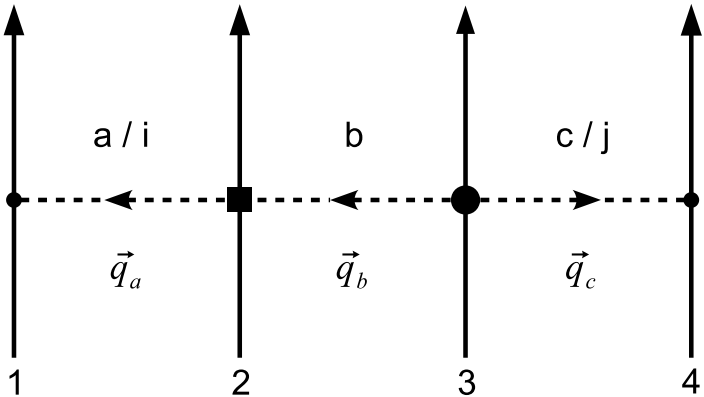}
\end{center}
\vspace{-11.3cm}
\caption{Generic form of the ''reducible'' chiral 4N-interaction. The square-box and 
filled circle symbolize effective single-nucleon vertices. The labels $a,b,c$ denote 
isospin-indices of exchanged pions and $i,j$ denote components of (spin)-vectors.}
\end{figure}

From the generic four-body diagram in Fig.\,1 the first-order "Hartree-Fock" 
contribution to the energy density $\rho \bar E(\rho)$  of nuclear matter is obtained 
by closing the four nucleon-lines in all possible ways. Inspecting the factorized 
expressions for the five reducible chiral $4N$-interactions written in 
eqs.(1,5,7,10,19) below, one sees that closing any nucleon-line to itself leads to a 
vanishing spin-trace or isospin-trace. In the case of four neutrons one has either 
a vanishing spin-trace or a zero from the totally anti-symmetric $\epsilon^{abc}
$-tensor, since $\pi^0$-exchange fixed $a=b=3$. As a consequence of this property, 
closed 4-ring and 3-ring diagrams vanish trivially and one needs to consider here 
only the closed 2-ring and 1-ring diagrams generated by the reducible chiral 
$4N$-interactions.
\begin{figure}[t!]
\begin{center}
\includegraphics[scale=0.6,clip]{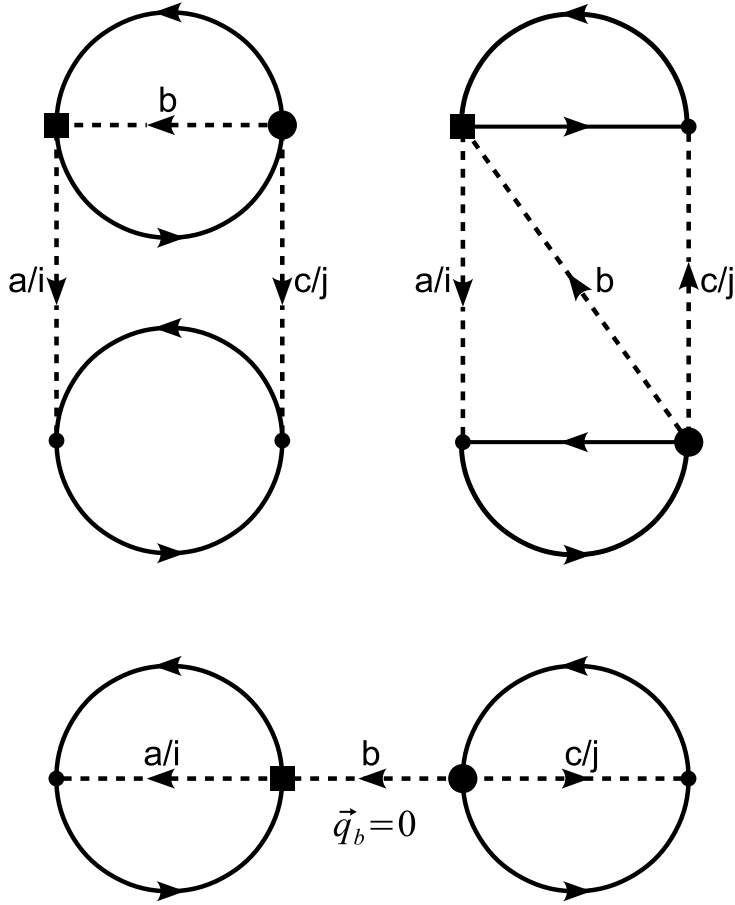}
\end{center}
\vspace{-3.cm}
\caption{Closed two-ring diagrams generated by the reducible chiral 4N-interaction. 
A symmetry factor $1/2$ occurs, if the  square-box and filled-circle vertex are 
identical. The diagram at the bottom vanishes due to the zero-momentum 
$\vec q_b = \vec 0$ of the exchanged central pion.}
\end{figure}
\begin{figure}[t!]
\begin{center}
\includegraphics[scale=0.6,clip]{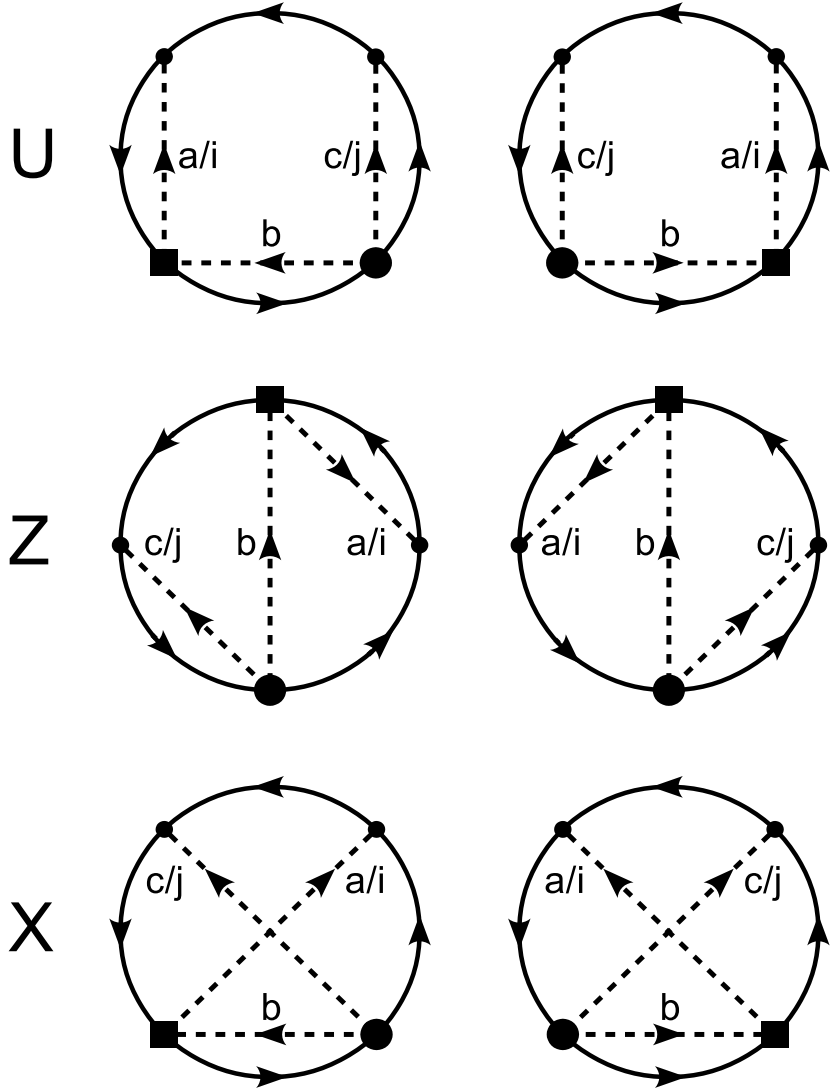}
\end{center}
\vspace{-2.cm}
\caption{Closed one-ring diagrams generated by the reducible chiral 4N-interaction. 
The diagrams in the first, second, and third row are referred to as U-type, Z-type, 
and X-type, respectively. A symmetry factor $1/2$ belongs to the Z-type diagrams, 
if the  square-box and filled-circle represent identical vertices.}
\end{figure}

The three possible 2-ring topologies are shown in Fig.\,2. One notices that the 
diagram at the bottom vanishes due to the zero-momentum $\vec q_b = \vec 0$ of the 
exchanged central pion. At least one of the adjoined vertices involves $\vec q_b$ 
linearly. Furthermore, the six possible 1-ring topologies are shown in Fig.\,3. 
According to the shape of the dashed (pion) line the diagrams in the first, second, 
and third row are referred to as U-type, Z-type, and X-type, respectively. Note 
also that if the square-box and filled-circle vertex represent identical vertices 
(this happens for class VII and I) a symmetry factor $1/2$ belongs to the 2-ring 
diagrams and to the Z-type 1-ring diagrams, and obviously identical copies of the 
U-type and X-type 1-ring diagrams need to be discarded. After these preparations we 
can now turn to the evaluation of the individual 1-ring and 2-ring diagrams 
generated by the five classes of reducible chiral $4N$-interactions.    
\subsection{Class VII}
We start (in reverse order) with class VII, which is quadratic in the spin-spin 
contact-coupling $C_T$. The corresponding factorized expression for this chiral
$4N$-interaction reads:  
\begin{equation}  V^{n} = -{C_T^2 g_A^2 \over f_\pi^2} \, \sigma_1^i \,
(\vec \sigma_2 \times \vec q_b)^i \, \tau_2^b \,{1\over(m_\pi^2+\vec q_b^{\,2})^2} 
\,(\vec \sigma_3\times \vec q_b)^j \, \tau_3^b \,\sigma_4^j\,.\end{equation} 
Note that there is an extra factor $2$ in comparison to eq.(3.60) in 
ref.\,\cite{evgeni4nb} due to the interchange symmetry $(12)\leftrightarrow(43)$ 
between pairs of nucleons. A straightforward  evaluation of the left 2-ring 
diagram in Fig.\,2 gives the following contribution from class VII to the energy per 
particle of isospin-symmetric nuclear matter:
\begin{eqnarray} \bar E(\rho) &=& {3C_T^2 g_A^2m_\pi^7 \over (2\pi^2 u)^3f_\pi^2}
\int_0^u\!\!dx\int_0^u\!\!dy \, x(u^2-y^2)(u-x)^2(2u+x) \bigg\{-8x y +6x \nonumber \\ 
&& \times \big[\arctan(2x+2y)- \arctan(2x-2y) \big] +(2x^2 -2y^2 -1)\ln{1+4(x+y)^2 
\over 1+4(x-y)^2}\bigg\}\,, \end{eqnarray}
with the dimensionless variable $u =k_f/m_\pi$. The nucleon density $\rho= 2k_f^3/3
\pi^2$ is related to the Fermi momentum $k_f$ in the usual way. In order to arrive 
at a double-integral representation we have used the reduction formula:
\begin{equation} \int_0^u\!\!dp\! \int_{-1}^1\!\!dz \, p^2 F\big(p z+\sqrt{u^2-p^2
(1-z^2)}\,\big) = \int_0^u\!\!dy \, (u^2-y^2) F(2y)\,, \end{equation} 
to eliminate a complicated angular integral, and a master formula for integrating $
F(|\vec p_1\!-\!\vec p_2|)$ over two Fermi spheres $|\vec p_{1,2}|<k_f$. The latter 
introduces the weighting-function $x^2(u-x)^2(2u+x)$. Actually, there exists an 
analytical solution of the double-integral in eq.(2), namely: 
\begin{eqnarray} \bar E(\rho) &=& {C_T^2g_A^2m_\pi^7 \over 140\pi^6f_\pi^2}
\Bigg\{ {u\over 160}-{31u^3\over 48}+{229u^5\over 20}-{179u^7\over 10} 
\nonumber \\ && +\bigg({3\over 16}+{343u^2\over 40}+20u^4-{142u^6\over 3}
\bigg)\arctan 2u\nonumber \\ && +\bigg({176u^6\over 3}-10u^4-{343u^2\over 
80}-{3 \over 32}\bigg) \arctan 4u \nonumber \\ && +\bigg(28u^5-{32u^7 
\over 5}+{35u^3 \over 8}-{7u\over 8}-{5 \over 128u} -{1 \over 3840 u^3}
\bigg) \ln(1+4u^2) \nonumber \\ && + \bigg({32u^7 \over 5}-28u^5-{35u^3 
\over 8}+{7u \over 32}+{5 \over 512u} +{1 \over 15360u^3} \bigg) 
\ln(1+16u^2) \Bigg\}\,. \end{eqnarray} 
Note that in the chiral limit $m_\pi \to 0$ only a term $u^7(128\ln 2-179)/10= 
-9.0277\,u^7$ remains. Let us also comment on the features of the other diagrams 
for class VII: The right 2-ring diagram in Fig\,2 leads to a vanishing isospin-trace 
in symmetric nuclear matter, whereas in pure neutron matter the left and right 
2-ring diagram cancel each other. The U-type and Z-type 1-ring diagrams in Fig.\,3 
cancel each other after taking the spin-trace, and the X-type 1-ring diagram leads 
to a vanishing spin-trace by itself.
\subsection{Class V}
Next we come to class V, which is induced by the Weinberg-Tomozawa $\pi\pi 
N\!N$-vertex in combination with the spin-spin contact-coupling $C_T$. The 
corresponding factorized expression for this chiral $4N$-interaction reads 
\cite{evgeni4nb}:  
\begin{equation}  V^l={C_T g_A^2\over 8f_\pi^4}\,\vec\sigma_1\cdot \vec q_a \,
\tau_1^a\,  {1\over m_\pi^2+\vec q_a^{\,2}} \, \epsilon^{abc} \, \tau_2^c \,
{1\over m_\pi^2+\vec q_b^{\,2}} \,(\vec \sigma_3\times \vec q_b)^j \, \tau_3^b
\,\sigma_4^j\,,\end{equation} 
and it obviously vanishes for four neutrons, where only $\pi^0$-exchange is 
possible ($\epsilon^{33c}=0$). The evaluation of the X-type 1-ring 
diagram in Fig.\,3 gives the following contribution from class V to the energy per 
particle of isospin-symmetric nuclear matter:
\begin{eqnarray} \bar E(\rho) &=& {3C_T g_A^2m_\pi^7 \over (2\pi)^6f_\pi^4u^3}
\int_0^u\!\!dx\int_0^u\!\!dy \, {x(u^2-y^2)\over 1+4x^2}(u-x)^2(2u+x) \bigg\{
-4x y \nonumber \\ && \times (1+4x^2+4y^2)+{1\over 4}\big[1+4(x+y)^2\big]
\big[1+4(x-y)^2\big] \ln{1+4(x+y)^2 \over 1+4(x-y)^2}\bigg\}\,. \end{eqnarray}
Note that the integral $\int_0^u\!dy\dots$ could be solved in terms of elementary 
functions ($\ln, \arctan$), but we refrain from presenting this rather lengthy 
expression. The same applies to the subsequent results given in eqs.(8,9,11,20,21).
Concerning the other diagrams for class V, one finds that both 2-ring diagrams in 
Fig.\,2 lead to a vanishing spin- and isospin-trace. Moreover, the U-type and Z-type 
1-ring diagrams in Fig.\,3  cancel each other after taking the spin-trace.

\subsection{Class IV}
Next we come to class VI, which is induced by the spin-spin contact-coupling $C_T$ in 
combination with a ''contraction'' of two ordinary $\pi N$-vertices. The 
corresponding factorized expression for this chiral $4N$-interaction reads 
\cite{evgeni4nb}: 
\begin{equation}  V^k = -{C_T g_A^4 \over 4f_\pi^4} \, \vec\sigma_1\cdot \vec q_a \,
\tau_1^a\,  {1\over m_\pi^2+\vec q_a^{\,2}} \, \big[\epsilon^{abc} \, \tau_2^c \,
\vec q_a\cdot\vec q_b+\delta^{ab}\,\vec\sigma_2\cdot(\vec q_a\times\vec q_b)\big]\,
{1\over (m_\pi^2+\vec q_b^{\,2})^2} \,(\vec \sigma_3\times \vec q_b)^j \, \tau_3^b
\,\sigma_4^j\,.\end{equation} 
Whereas the left 2-ring diagram has a vanishing isospin-trace, one obtains from the 
right 2-ring diagram the following contribution to the energy per particle 
of isospin-symmetric nuclear matter:
\begin{eqnarray} \bar E(\rho) &=& {3C_T g_A^4m_\pi^7 \over 16\pi^6f_\pi^4u^3}
\int_0^u\!\!dx\int_0^u\!\!dy \, {x(u^2-y^2)\over 1+4x^2}(u-x)^2(2u+x) \nonumber 
\\ && \times \bigg\{x y(1-12x^2+4y^2) +8x^3\big[\arctan(2x+2y)- \arctan(2x-2y)
\big] \nonumber \\ && +\Big(3x^4-{3x^2 +y^2 \over 2}-2x^2y^2-y^4-{1 \over 
16} \Big) \ln{1+4(x+y)^2 \over 1+4(x-y)^2} \bigg\} \,. \end{eqnarray}
By rewriting the spin- and isospin-factors as a product of two anti-commutators one 
finds that the U-type and Z-type 1-ring diagrams add to zero. On the other hand, the
X-type 1-ring diagrams in Fig.\,3 give a contribution to $\bar E(\rho)$ of the form:
\begin{eqnarray} \bar E(\rho) &=& {3C_T g_A^4m_\pi^7 \over 16\pi^6f_\pi^4u^3}
\int_0^u\!\!dx\int_0^u\!\!dy \, {x(u^2-y^2)\over 1+4x^2}(u-x)^2(2u+x) \nonumber 
\\ && \times \bigg\{ 2x y(1+4x^2+4y^2) +8x^3\big[\arctan(2x-2y)-\arctan(2x+2y)
\big] \nonumber \\ &&  +\Big(x^2-2x^4 +4x^2y^2-y^2-2y^4-{1\over 8}\Big)
\ln{1+4(x+y)^2 \over 1+4(x-y)^2} \bigg\}\,. \end{eqnarray}
In the case of pure neutron matter the left and right 2-ring diagram cancel each 
other, likewise do this the U-type and Z-type 1-ring diagrams,  and the X-type 1-ring 
diagrams vanish by themselves (because for four neutrons the $\epsilon^{abc}$-term 
in $V^k$ is absent).

\subsection{Class II}
Next we come to class II, which is induced by the Weinberg-Tomozawa $\pi\pi 
N\!N$-vertex in combination with the ''contraction''of two ordinary $\pi 
N$-vertices. The corresponding factorized expression for this $3\pi$-exchange 
$4N$-interaction reads 
\cite{evgeni4nb}: 
\begin{eqnarray}  V^c &=& -{g_A^4 \over 32f_\pi^6} \, \vec\sigma_1\cdot \vec q_a \,
\tau_1^a\,  {1\over m_\pi^2+\vec q_a^{\,2}} \, \epsilon^{abd}\, \tau_2^d \, {1\over 
m_\pi^2+\vec q_b^{\,2}} \, \nonumber \\ && \times \big[\epsilon^{bce} \, \tau_3^e \,
\vec q_b\cdot\vec q_c +\delta^{bc}\,\vec\sigma_3\cdot(\vec q_b\times\vec q_c)
\big]\,{1\over m_\pi^2+\vec q_c^{\,2}} \,\vec \sigma_4\cdot \vec q_c \, \tau_4^c\,,
\end{eqnarray} 
and it obviously vanishes for four neutrons, since $\epsilon^{33d}=0$. The evaluation 
of the left 2-ring diagram in Fig.\,2 gives the following contribution to the energy 
per particle of isospin-symmetric nuclear matter:
\begin{eqnarray} \bar E(\rho) &=& {3g_A^4m_\pi^7 \over (2\pi f_\pi)^6u^3}
\int_0^u\!\!dx\int_0^u\!\!dy \, {x^3(u^2-y^2)\over (1+4x^2)^2}(u-x)^2(2u+x) 
\bigg\{ 4x y\nonumber \\ && \times (1+4x^2+4y^2)-{1\over 4}\big[1+4(x+y)^2
\big] \big[1+4(x-y)^2\big] \ln{1+4(x+y)^2 \over 1+4(x-y)^2} \bigg\}\,, 
\end{eqnarray}
whereas the right 2-ring diagram produces a vanishing spin-trace. In the case of
the U-type 1-ring diagrams in Fig.\,3, three (out of four) Fermi sphere integrals 
over the pion-propagators and momentum-dependent interactions factorize by making use 
of tensor-contractions. For this reason one can represent the corresponding 
contribution to  $\bar E(\rho)$  as a one-parameter integral of the form:
\begin{equation} \bar E(\rho) = -{3g_A^4m_\pi^7 \over 4(4\pi f_\pi)^6u^3}
\int_0^u\!\!dx \, {G_V^2(x) \over x} \Big[2G_S(x)+G_T(x)\Big]\,. \end{equation}
Here, we have introduced the auxiliary functions: 
\begin{equation} G_V(x)= u(1+u^2+x^2)-{1\over 4x}\big[1+(u+x)^2\big]\big[
1+(u-x)^2\big]  \ln{1+(u+x)^2 \over 1+(u-x)^2}\,, \end{equation}
\begin{eqnarray} G_S(x)&=& {4u x\over 3}(2u^2 -3)+4x \big[ 
\arctan(u+x) + \arctan(u-x)\big]\nonumber \\ &&  +(x^2-u^2-1) \ln{1+(u+x)^2
\over 1+(u-x)^2} \,, \end{eqnarray}
\begin{eqnarray}G_T(x)&=& {u x \over 6}(8u^2 +3x^2)-{u \over 
2x}(1+u^2)^2+{1 \over 8}\bigg[{(1+u^2)^3 \over x^2}\nonumber \\ && -x^4  
+(1-3u^2)(1+u^2-x^2)\bigg]\ln{1+(u+x)^2 \over 1+(u-x)^2} \,, \end{eqnarray}
defined by vectorial and tensorial Fermi sphere integrals over a pion-propagator
(see also eq.(A.1) in ref.\,\cite{4body}). In the case of the Z-type 1-ring diagrams
in Fig.\,3, the two Fermi sphere integrals over the pion-propagators associated to 
the short pion-lines factorize via tensors. For the remaining two Fermi sphere 
integrals the angular part can be carried out, and in this procedure the third 
pion-propagator introduces the logarithmic function: 
\begin{equation} {\bf L}= \ln{1+(x+y)^2 \over 1+(x-y)^2} \,. \end{equation}
Putting all pieces together, one obtains the following double-integral representation 
for the contribution of the Z-type 1-ring diagrams from class II to the energy per 
particle of nuclear matter:
\begin{eqnarray} \bar E(\rho) &=& {3g_A^4m_\pi^7 \over (4\pi f_\pi)^6u^3}
\int_0^u\!\!dx\int_0^u\!\!dy \, G_V(x) \bigg\{G_S(y) \bigg[2y+{{\bf L} 
\over 2x}(x^2-y^2-1)\bigg] \nonumber \\ &&  + {G_T(y) \over 4y} \bigg[y^2-
3x^2-3+{{\bf L}\over 4xy} \Big(3(1+x^2)^2+2y^2-2x^2y^2-y^4\Big)\bigg]\bigg\}
\,. \end{eqnarray}
In the case of the X-type 1-ring diagrams in Fig.\,3 the assignment of momenta to 
the virtual pions is now such that the previous factorization approach does not 
work anymore. Only the Fermi sphere integral associated to the nucleon-line between 
the square-box and filled-circle vertex can be performed analytically. Hence, the 
contribution of the X-type 1-ring diagrams from class II to the energy per particle 
of nuclear matter result in the form:
\begin{equation}  \bar E(\rho) = {9g_A^4m_\pi^7 \over (4\pi f_\pi)^6 u^3}
\int\limits_{|\vec p_j|<u}\!\!\!{d^3p_1\, d^3p_2\,d^3p_3 \over (2\pi)^3} \, { G_V(|\vec 
\eta\, |)\over (1+ \vec q_1^{\,2}) (1+ \vec q_2^{\,2}) \, \vec \eta^{\,2}}
\, [(\vec q_1- \vec q_2)\cdot\vec q_2] \,\, (\vec q_1\cdot\vec \eta\,)
\,,  \end{equation}
with $\vec q_1 = \vec p_1-\vec p_3$, $\vec q_2 = \vec p_2-\vec p_3$ and 
$\vec \eta=\vec p_1+\vec p_2-\vec p_3$, where all momentum-vectors have been divided 
by the pion mass $m_\pi$. Note that due to rotational invariance it is sufficient to 
parametrize the nine-dimensional integral in eq.(18) by three radii ($0<p_{1,2,3}<u$), 
two directional cosines ($-1<z_{1,2}<1$), and one azimuthal angle ($0<\varphi<2\pi$).
   
\subsection{Class I}
Finally we come to class I, which is induced by a pair of ''contractions'' of two 
ordinary $\pi N$-vertices. The corresponding factorized expression for this 
$3\pi$-exchange $4N$-interaction reads: 
\begin{eqnarray}  V^a &=& {g_A^6 \over 16f_\pi^6} \, \vec\sigma_1\cdot \vec q_a \,
\tau_1^a\,  {1\over m_\pi^2+\vec q_a^{\,2}} \, \big[\epsilon^{abd} \, \tau_2^d \,
\vec q_a\cdot\vec q_b +\delta^{ab}\,\vec\sigma_2\cdot(\vec q_a\times\vec q_b)\big]\,
\nonumber \\ && \times {1\over (m_\pi^2+\vec q_b^{\,2})^2} \,\big[\epsilon^{bce} \, 
\tau_3^e \,\vec q_b\cdot\vec q_c +\delta^{bc}\,\vec\sigma_3\cdot(\vec q_b\times
\vec q_c)\big]\,{1\over m_\pi^2+\vec q_c^{\,2}} \,\vec \sigma_4\cdot \vec q_c \, 
\tau_4^c\,, \end{eqnarray}
Note that there is again an extra factor $2$ in comparison to eq.(3.33) in 
ref.\,\cite{evgeni4nb} due to the interchange symmetry $(12)\leftrightarrow(43)$ 
between pairs of nucleons. The evaluation of the left 2-ring diagram in Fig.\,2 
gives the following contributions to the energy per particle of isospin-symmetric 
nuclear matter:
\begin{eqnarray} \bar E(\rho) &=& {3g_A^6m_\pi^7 \over (2\pi f_\pi)^6u^3}
\int_0^u\!\!dx\int_0^u\!\!dy \,{x^3(u^2-y^2)\over (1+4x^2)^2}(u-x)^2 (2u+x)
\nonumber \\ && \times \bigg\{-2x y(1+20x^2+4y^2)+32x^3\big[\arctan(2x+2y)-
\arctan(2x-2y)\big]\nonumber \\ && +\Big(10x^4-5x^2+y^2-12x^2y^2+2y^4+
{1\over 8}\Big) \ln{1+4(x+y)^2\over 1+4(x-y)^2}\bigg\}\,, \end{eqnarray}
and to the energy per particle of pure neutron matter:
\begin{eqnarray} \bar E_n(\rho_n) &=& {g_A^6m_\pi^7 \over (2\pi f_\pi)^6u^3}
\int_0^u\!\!dx\int_0^u\!\!dy \,{x^3(u^2-y^2)\over (1+4x^2)^2}(u-x)^2 (2u+x)
\nonumber \\ && \times \bigg\{x y(1-12x^2+4y^2)+8x^3\big[\arctan(2x+2y)-
\arctan(2x-2y)\big]\nonumber \\ && +\Big(3x^4-2x^2y^2-y^4-{3x^2+y^2 
\over 2}-{1 \over 16}\Big) \ln{1+4(x+y)^2 \over 1+4(x-y)^2}\bigg\}\,. 
\end{eqnarray} 
The neutron density $\rho_n$ is related to the neutron Fermi momentum $k_n$ by 
$\rho_n=k_n^3/3\pi^2$. In eq.(21) and all following formulas for $\bar E_n(\rho_n)$ 
the dimensionless variable $u$ has the meaning $u  =k_n/m_\pi$. Next one notices 
that for class I the right 2-ring diagram in Fig.\,2 produces vanishing spin-trace. 

Furthermore, one obtains from the U-type 1-ring diagram in Fig.\,3 the following 
contributions:
\begin{equation} \bar E(\rho) = {g_A^6m_\pi^7 \over 4(4\pi f_\pi)^6u^3}
\int_0^u\!\!dx\,\bigg\{ \Big[8G_S^2(x)+G_T^2(x)\Big]K_S(x)+ \Big[8G_S(x)
+G_T(x)\Big]G_T(x) K_T(x) \bigg\} \,, \end{equation}
\begin{equation} \bar E_n(\rho_n) = {g_A^6m_\pi^7 \over 12(4\pi f_\pi)^6u^3}
\int_0^u\!\!dx \,\bigg\{ \Big[2G_S^2(x)+G_T^2(x)\Big]K_S(x)+ \Big[G_T(x)-4G_S(x)
\Big]G_T(x) K_T(x) \bigg\} \,, \end{equation}
where we have introduced two new auxiliary functions:
\begin{equation} K_S(x)= 2u-3\arctan(u+x)-3\arctan(u-x) + 
 {2+u^2-x^2\over 2x}  \ln{1+(u+x)^2 \over 1+(u-x)^2} \,,\end{equation} 
\begin{equation} K_T(x)= {u\over 4x^2}(3+3u^2-x^2)+{1\over 16 x^3}\Big[
x^2(x^2+2u^2-2)-3(1+u^2)^2\Big]\ln{1+(u+x)^2 \over 1+(u-x)^2}\,, \end{equation}
defined by a tensorial Fermi sphere integral over a squared pion-propagator.

On the other hand the Z-type 1-ring diagrams give the following contributions:
\begin{eqnarray} \bar E(\rho) &=& {g_A^6m_\pi^7 \over (4\pi f_\pi)^6u^3}
\int_0^u\!\!dx\int_0^u\!\!dy\, \bigg\{2 G_S(x)G_S(y)\,[{\bf D}-{\bf L}]+G_S(x) G_T(y) 
\nonumber \\ && \times \bigg[2{\bf D}-{6x \over y}+{{\bf L}\over 2y^2}(3+3x^2-y^2)
\bigg]+ G_T(x) G_T(y) \nonumber \\ && \times \bigg[{{\bf D}\over 2} +{3y\over 8x}
+{27\over 16x y} -{{\bf L}\over 64 x^2y^2}(27+60y^2+2x^2y^2+6y^4)\bigg]\bigg\} \,,
\end{eqnarray} 
\begin{eqnarray} \bar E_n(\rho_n) &=& {g_A^6m_\pi^7 \over 6(4\pi f_\pi)^6u^3}
\int_0^u\!\!dx\int_0^u\!\!dy \,\bigg\{G_S(x)G_S(y)\, [{\bf D}-{\bf L}] +G_S(x) G_T(y) 
\nonumber \\ && \times \bigg[{6x \over y}-2{\bf D} +{{\bf L}\over 2y^2}(y^2-3x^2-3)
\bigg]+G_T(x)G_T(y) \nonumber \\ && \times\bigg[{\bf D}+{3y\over 4x}+{27\over 8x y} 
-{{\bf L}\over 32 x^2y^2}(27+60y^2+2x^2y^2+6y^4)\bigg]\bigg\} \,,\end{eqnarray} 
with the rational function:
\begin{equation}{\bf D} = {1 \over 1+(x-y)^2}-{1 \over 1+(x+y)^2}\,, \end{equation}
arising from an angular integral $\int_{-1}^1\!dz\dots$ over a squared 
pion-propagator.

Finally, the X-type 1-ring diagram leads to contributions of the form:
\begin{eqnarray}  \bar E(\rho) &=& {3g_A^6m_\pi^7 \over (4\pi f_\pi)^6 u^3}
\int\limits_{|\vec p_j|<u}\!\!\!{d^3p_1\, d^3p_2\,d^3p_3\over(2\pi)^3}\,{1\over(1+ 
\vec q_1^{\,2}) (1+ \vec q_2^{\,2})}\nonumber \\ && \times\bigg\{ 
3\vec q_1^{\,2}\,\vec q_2^{\,2} \Big[K_S(|\vec \eta\,|) -2 K_T(|\vec \eta\,|)\Big] +
(\vec q_1\cdot\vec q_2)^2 \Big[7K_T(|\vec \eta\,|)-K_S(|\vec \eta\,|)\Big]
\nonumber \\ && +\,{3 \over \vec \eta^{\,2}} K_T(|\vec \eta\,|) \Big[  6\,\vec q_1^{\,2}
\,(\vec \eta\cdot\vec q_2)^2- 7(\vec q_1\cdot\vec q_2)\,(\vec \eta\cdot \vec q_1)
\, (\vec \eta\cdot\vec q_2)\Big] \bigg\}\,, \end{eqnarray}
\begin{eqnarray}  \bar E_n(\rho_n) &=& {g_A^6m_\pi^7 \over (4\pi f_\pi)^6 u^3}
\int\limits_{|\vec p_j|<u}\!\!\!{d^3p_1\, d^3p_2\,d^3p_3\over(2\pi)^3}\,{1\over(1+\vec 
q_1^{\,2}) (1+ \vec q_2^{\,2})}\nonumber \\ && \times \bigg\{ 
3(\vec q_1\cdot \vec q_2)^2\Big[K_S(|\vec \eta\,|) +K_T(|\vec \eta\,|)\Big] 
- \vec q_1^{\,2}\,\vec q_2^{\,2} \Big[K_S(|\vec \eta\,|)+2K_T(|\vec \eta\,|)\Big]
\nonumber \\ && +\,{3 \over \vec \eta^{\,2}} K_T(|\vec \eta\,|) \Big[  
2\,\vec q_1^{\,2}\,(\vec \eta\cdot \vec q_2)^2- 3(\vec q_1\cdot\vec q_2)\,(\vec 
\eta\cdot\vec q_1) \, (\vec \eta\cdot\vec q_2)\Big] \bigg\}\,, \end{eqnarray}
with $\vec q_1 = \vec p_1-\vec p_3$, $\vec q_2 = \vec p_2-\vec p_3$ and $\vec \eta= 
\vec p_1+\vec p_2-\vec p_3$. 

This completes our presentation of semi-analytical results for the energy per 
particle of isospin-symmetric nuclear matter $\bar E(\rho)$ and pure neutron matter 
 $\bar E(\rho)$ as derived from the reducible chiral $4N$-interactions. The given 
formulas allow for an easy and accurate numerical evaluation  \cite{code} of these 
leading-order chiral four-body correlations.  

\section{Results and discussion}
We are now in the position to present numerical results. The physical parameters 
related to pion-exchange are: $g_A = 1.29$ (nucleon axial-vector coupling constant), 
$f_\pi = 92.4$\,MeV (pion decay constant), and $m_\pi = 135\,$MeV (neutral pion mass).
For the spin-spin contact-coupling $C_T$ we choose two opposite values, $C_T=
0.22\,$fm$^2$ and $C_T=-0.45\,$fm$^2$, taken from table I in ref.\cite{darmstadt}.

The contributions to the energy per particle $\bar E(\rho)$ of isospin-symmetric 
nuclear matter are presented in Fig.\,4 for the small $C_T$-value, $C_T=0.22\,
$fm$^2$, in the density region $\rho_0/4<\rho <2 \rho_0$, with $\rho_0=0.16\,
$fm$^{-3}$ the empirical saturation density. The dashed lines (with appropriate 
labels) show separately the five classes (VII, V, IV, II, I) and the full line gives 
their total sum. One observes that class VII (proportional to $C_T^2$) is smallest 
in magnitude and that other classes (V+IV and II+I) cancel each other to a large 
extent. The largest repulsive and attractive contributions are provided by the 
$C_T$-independent classes II and I, respectively. As a consequence of these 
compensations the net attraction stays above $-1.3\,$MeV for all densities 
$\rho<0.36\,$fm$^{-3}$. The analogous results for the larger negative $C_T$-value, 
$C_T=-0.45\,$fm$^2$, are shown in Fig.\,5. One notices that class VII ($\sim C_T^2$) 
is still smallest in magnitude and that the classes V and IV have changed their 
position in comparison to the sequence in Fig.\,4. The net attraction reaches now a 
value of $-1.7\,$MeV at $\rho=0.36\,$fm$^{-3}$. Exploring the whole range $|C_T|
<0.5\,$fm$^2$ of the spin-spin contact-coupling $C_T$, one finds that the 
$4N$-attraction in nuclear matter does not exceed values of $-1.3\,$MeV for 
densities $\rho <2 \rho_0$.

\begin{figure}[t!]
\begin{center}
\includegraphics[scale=0.5,clip]{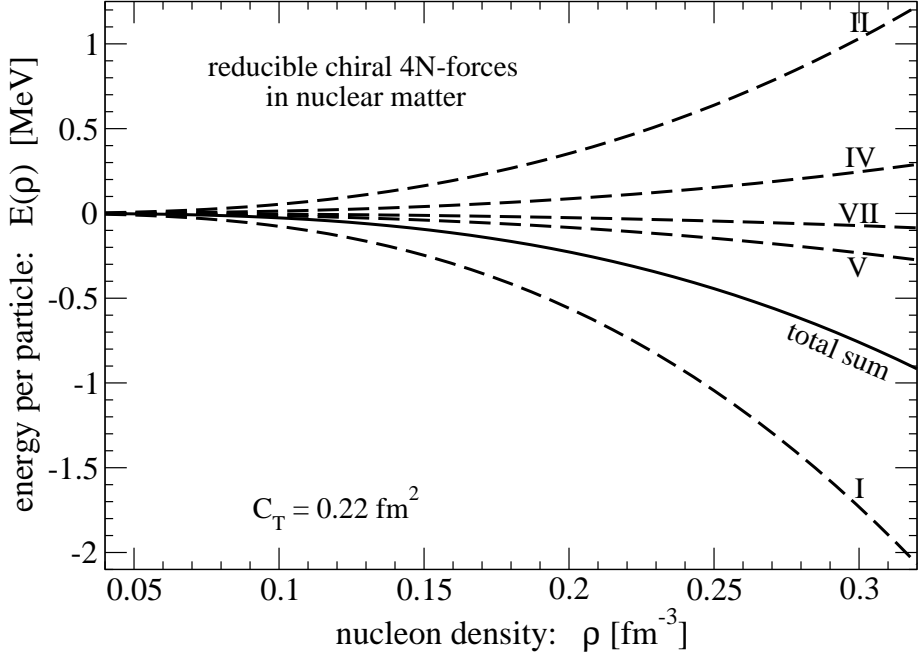}
\end{center}
\vspace{-.6cm}
\caption{Reducible chiral $4N$-interactions in isospin-symmetric nuclear matter. 
The dashed lines show the individual contributions from the five classes for 
$C_T=0.22$\,fm$^2$, and the full line gives their total sum.}
\end{figure}

\begin{figure}[t!]
\begin{center}
\includegraphics[scale=0.5,clip]{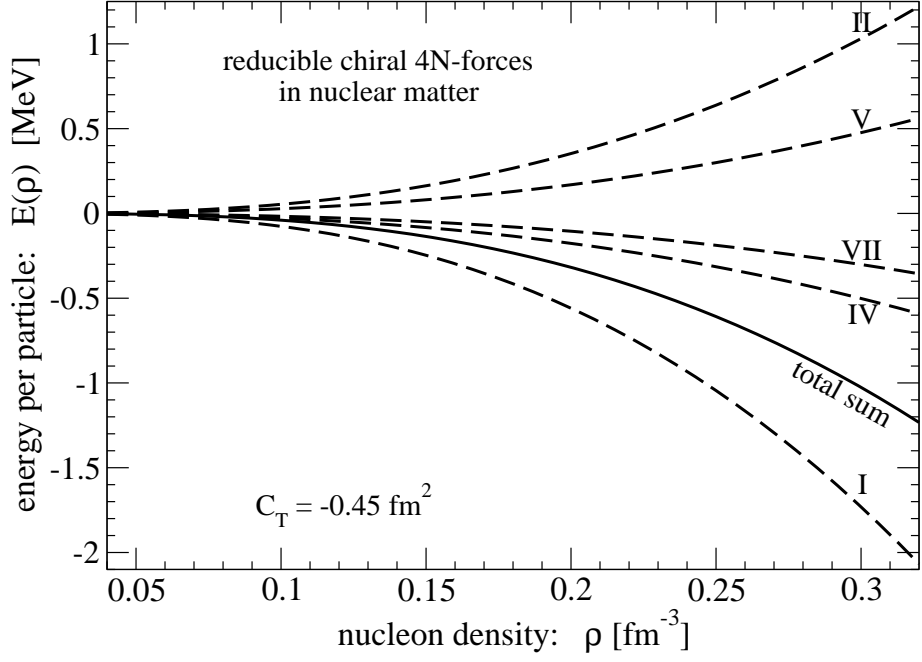}
\end{center}
\vspace{-.6cm}
\caption{Reducible chiral $4N$-interactions in isospin-symmetric nuclear matter. 
The dashed lines show the individual contributions from the five classes for 
$C_T=-0.45$\,fm$^2$, and the full line gives their total sum.}
\end{figure}

\begin{figure}[h!]
\begin{center}
\includegraphics[scale=0.5,clip]{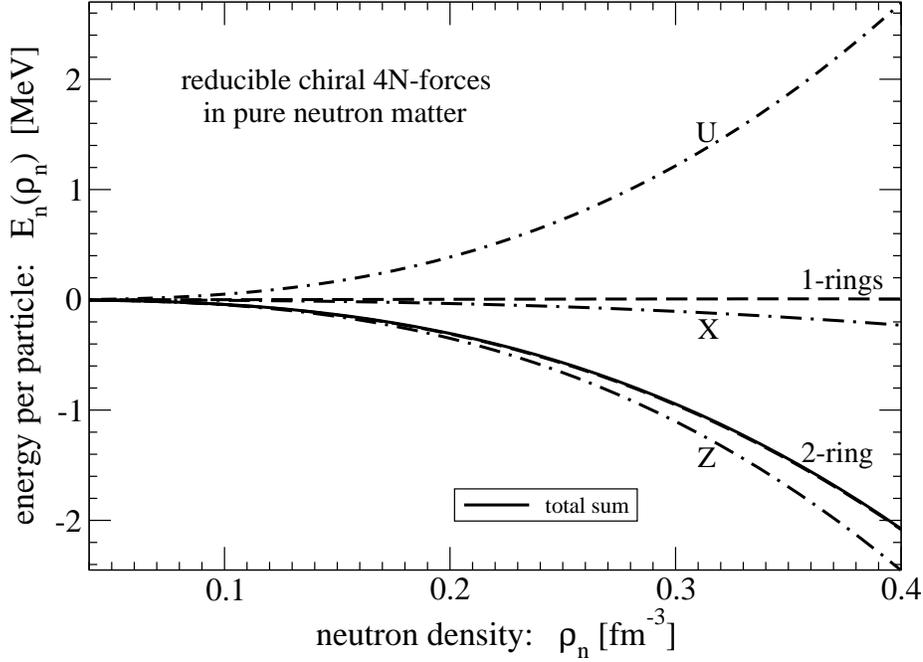}
\end{center}
\vspace{-.6cm}
\caption{Reducible chiral $4N$-interactions in pure neutron matter. The
conributions from 2-ring and 1-ring diagrams are shown separately. The U-type, 
Z-type, and X-type diagrams cancel each other almost completely.}
\end{figure}

In the case of pure neutron matter only class I (see subsection 2.5) contributes and 
the result for the energy per particle $\bar E_n(\rho_n)$ comes out parameterfree.
Fig.\,6 shows by the dashed lines the contributions from the 2-ring diagram and the 
sum of the three 1-ring diagrams in the density region $0.04\,$fm$^{-3}<\rho_n<0.4
\,$fm$^{-3}$. One recognizes that the latter part is extremely small. This feature 
has its origin in the almost complete compensation between the contributions from 
the U-type, Z-type and X-type 1-ring diagrams, which are shown by the three 
dashed-dotted lines in Fig.\,6. Interestingly, the net result for  $\bar E_n(\rho_n)$ 
is thus entirely determined by the 2-ring part written in eq.(21). The corresponding 
attractive energy per particle goes approximately as $\rho_n^{7/3}$ and it reaches 
the value of about $-2.1$\,MeV at the (relatively high) neutron density of $\rho_n=
0.4$\,fm$^{-3}$. 

In passing we note that our numerical results for the chiral four-body correlations 
in nuclear and neutron matter agree perfectly \cite{thomas} with those of the 
Darmstadt group  \cite{darmstadt} if the (unessential) regulator-function $f_R$ is 
set to $f_R=1$.  

In summary one can conclude that the chiral four-nucleon correlations studied in 
this work are at least one order of magnitude smaller than those provided by the 
strongly coupled $\pi N\Delta$-system \cite{4body} with its small mass-gap of  
$293$\,MeV\,$\simeq 1.1 k_{f0}$.  

\section*{Acknowledgements}
We thank E. Epelbaum, T. Kr\"uger and A. Schwenk for informative discussions.


\begin{thebibliography}{99}
\bibitem{code} A Mathematica code can be obtained from the first author upon 
request.\vs
\end{thebibliography}
\end{document}